\documentstyle[multicol, eqsecnum, prd, aps]{revtex}

\def\beginwide{
        \end{multicols} \vspace*{-0.5cm} \noindent
        \rule{3.5in}{.1mm}\rule{.1mm}{5mm} \widetext \medskip }
\def\endwide{
        \hspace*{3.5in}~\rule[-5mm]{.1mm}{5mm}\rule{3.5in}{.1mm}
        \begin{multicols}{2}\narrowtext \vspace*{-1.0cm} \noindent }
\def\beginwidetop{
        \end{multicols} \vspace*{-0.5cm} \noindent
        \widetext \medskip }
\def\endwidebottom{
        \begin{multicols}{2} \vspace*{-1.0cm} \noindent }
\begin{document}
\draft
\title{Volume elements of spacetime and\\a quartet of scalar fields}
\author{Frank Gronwald$^1$, Uwe Muench$^1$, Alfredo
  Mac\'{\i}as$^2$, Friedrich W.\ 
  Hehl$^1$}
\address{$^1$ Institute for Theoretical Physics, University of Cologne,
  D-50923 K{\"o}ln, Germany\\
$^2$ Departamento de
  F\'{\i}sica, Universidad Aut\'onoma Metropolitana-Iztapalapa,\\
  P.O. Box 55-534, 09340 M\'exico D.F., M\'exico} 
\date{received ...}
\maketitle
\begin{abstract}
    Starting with a `bare'
    4-dimensional differential manifold as a model of spacetime, we
    discuss the options one has for defining a volume element which
    can be used for physical theories. We show that one has to
    prescribe a scalar density $\bbox{\sigma}$. Whereas conventionally
    $\sqrt{\left| \det g_{ij}\right|}$ is used for that purpose, with
    $g_{ij}$ as the components of the metric, we point out other
    possibilities, namely $\bbox{\sigma}$ as a `dilaton' field
    or as a derived quantity from either a linear connection or a
    quartet of scalar fields, as suggested by Guendelman and Kaganovich.
    \hfill {\em file volume5a.tex, 1997-12-15}
\end{abstract}

\begin{multicols}{2}\narrowtext 

\section{Introduction}
A fundamental premise is that gravity is intimately intertwined with the 
geometry of spacetime. At the classical level, general relativity 
captures this idea in an elegant way. The identification of the 
gravitational field with the curvature of spacetime has led to the most 
dramatic predictions of general relativity, such as the existence of 
black holes and the occurrence of the big bang. This leads to the 
viewpoint that a primary goal of any quantum theory should be the quantum 
structure of spacetime.

There are two programs which have made great strides forward into the
quantum gravity in the last twelve years, namely, one perturbative, i.e.\ 
{\em the quantum superstring}, and one nonperturbative,
i.e.\ {\em Dirac's canonical quantization} program, applying Ashtekar
realization of Einsteins's general relativity and the loop representation
for quantum mechanics.

At the first sight, it would appear that the two approaches are 
incompatible. Ashtekar's program applies pure Einstein gravity, without 
involving anything else, whereas the superstring is a 
{\em theory of everything} and, because of unitarity, would involve all 
interactions in its perturbative expansion. These differences will appear at
Planck energies only, and only around these energies one might be able to
decide between string theory and canonical gravity.

On the other hand, the study of gravitational interactions coupled to 
Maxwell and scalar dilaton fields has been the subject of recent 
investigations. Dilaton fields appear naturally in the low energy limit
of string theory. Therefore, the study of the scalar fields is of 
importance for the understanding of more general theories. Moreover, it has
been demostrated that the metric-affine gravity theories contain the 
axi-dilatonic sector of low energy string theory. The gravitational 
interactions involving the axion and dilaton may be derived from a 
geometrical action principle involving the curvature scalar with a 
non-Riemannian connection, i.e.\ the axi-dilatonic sector of the low 
energy string theory can be expressed in terms of geometry with torsion and
nonmetricity \cite{stelle1}. 

The local geometry of spacetime is usually characterized by two
independent concepts: The concept of a linear connection (parallel
transport) and the concept of a metric (length and angle
measurements). Both, the linear connection and the metric are physical
fields which have to be determined from field equations of the
gravitational theory under consideration.

In physics we do not only want to postulate the fields that are needed
in order to formulate a theory. We also find it desirable to explain
the existence of these fields by means of some fundamental principle.
Therefore we are led to wonder what fundamental principles would
suggest the existence of the linear connection and the metric,
respectively.

The existence of the {\em linear connection} is quite satisfactorily
explained by a symmetry principle which is believed to be fundamental:
This is the principle of local gauge invariance. In the case of
gravity we focus on {\em external} symmetries, i.e., symmetry
transformations of spacetime. Local gauge invariance then reflects the
invariance of a physical system under such transformations. As in the
Yang-Mills theories, this 
requires the introduction of a gauge connection which, in turn, allows
to define parallel transport. The actual gauge connection depends on
the specific symmetry under consideration and is, in a gauge approach
to gravity, of the form of a linear connection together
with the coframe; for details, see \cite{hehl95,gron97}.

In contrast to this, it is not clear how to derive the {\em metric}
{}from some fundamental principle. Usually the existence of the metric
is simply assumed, sometimes in disguise of a local symmetry group
which contains an orthogonal subgroup. Therefore it is quite natural
to ask whether the metric itself is a fundamental quantity, a derived
quantity, or a quantity which can be substituted by some more
fundamental field. To investigate this question is one of the
motivations for this article. Another motivation is our desire to
understand the physical consequences of the resulting theories and
whether or not they are more suitable for quantization.
As we will see, both motivations will lead to interrelated
questions and structures.

Our starting point is the observation that the metric is commonly
taken to define a volume element in order to be able to perform
integrations on spacetime. Integrated objects are clearly of
fundamental importance in physics. However, the definition of a volume
element on spacetime is also possible without reference to any metric.
This general subject will be explained in
Sec.\ \ref{sec:sec2}. Basically, a volume 
element can be defined on any differentiable manifold as the
determinant of a parallelepiped defined in terms of $n$ vectors, if
$n$ is the dimension of the manifold. Then no {\em absolute} volume
measure exists. However, {\em proportions} of different volumes can be
determined. More explicitly, one finds that the volume element is the
Levi-Civita $n$-form density transvected with the components of the
$n$ linearly independent vectors spanning the parallelepiped.  Such a
volume is an (odd) density of weight $-1$.  In order to define an
integral, we need then additionally a scalar density of weight $+1$.

Usual physical fields are no densities. Therefore the common practice
is to take the metric and to build a density according to
$\sqrt{|{\det}g_{ij}|}$. But there exist alternatives, as we will
point out in Sec.\ \ref{sec:sec3}. They open the gate to alternative
theories of gravitation.

Subsequently, in Sec.\ \ref{sec:sec5}, we will follow up one
possibility, namely the quartet of scalar fields, as proposed in
\cite{Guendel996a,Guendel997a,Guendel997b}. In an appendix, we will
provide some mathematical background for the differentiation of some
quantities closely related to the volume element.

\section{Integration on spacetime}
We\label{sec:sec2} model spacetime as a 4-dimensional differentiable
manifold, which is assumed to be paracompact, Hausdorff, and
connected. We will restrict ourselves to four dimensions.
Generalization to arbitrary dimensions is straightforward. In order
to be able to formulate physical laws on such a spacetime, we have to
come up with suitably defined integrals. If we want, for example, to
specify a {\em scalar} action functional $W$ of a physical system,
\begin{equation} W=\int L = \int\widetilde{\epsilon}\;\widehat{L} \;,
\label{eq:action}\end{equation}
then, taking the integral in its conventional (Lebesgue) meaning, the
Lagrangian $L$ has to be an odd 4-form in order to make the
integral (\ref{eq:action}) really a scalar. Incidentally, a $p$-form
$\omega = \frac{1}{p!}\, \omega_{i_1\dots i_p}\,dx^{i_1}\wedge\dots\wedge
dx^{i_p}$ is called {\em even} if it is
invariant under a diffeomorphism $x^i\rightarrow {x'}^i(x^j)$ with
$\text{det}\bigl(\partial x^j/\partial{x'}^i\bigr) < 0.$ It is called
{\em odd} if it changes its sign under such a diffeomorphism; for
even and odd forms, see Burke \cite{burk85}, Bott and Tu \cite{bott82},
and also \cite{hehl95}. 

Now, any odd 4-form can be split into a product of a 0-form (or
scalar) $\widehat{L}$ and another odd 4-form $\widetilde{\epsilon}$.  We
assume that we did the splitting in such a way that the properties of
the physical system are subsumed in the scalar $\widehat{L}$ and the
properties of spacetime in the {\em odd volume 4-form}
$\widetilde{\epsilon}$. In the case of {\em gravity} such a distinction may
be ambiguous, because the properties of spacetime themselves are parts
of the physical system.

Let us consider a trivial physical system with $\widehat{L}=1$. Then the
integral measures the volume of the corresponding piece of spacetime,
\begin{equation} \text{Vol}=\int\widetilde{\epsilon} \;.
\label{volume}\end{equation}
For that reason $\widetilde{\epsilon}$ is called a volume form or, more
colloquially, a {\em volume element} of spacetime. This quantity can
be split again into two pieces. 

As the first piece we have the Levi-Civita $\bbox{\epsilon}$ in
mind which is a very special geometric object. The Levi-Civita
$\bbox{\epsilon}$ can be defined on any differential manifold.
It is the `purely geometrical' volume element (see Grassmann's theory
of extension [{\em Ausdehnungslehre}] or the discussion of Laurent
\cite{Laurent}). In order to define $\bbox{\epsilon}$, we recall
that, besides the components of the Kronecker symbol $\delta^j_i$, the
components $\bbox{\epsilon}_{ijkl}$ of the Levi-Civita
$\bbox{\epsilon}$ are, by assumption, {\em numerically
  invariant} under diffeomorphisms.  There exist no other quantities
of this kind, apart from $\bbox{\epsilon}^{ijkl}$ of
Sec.\ \ref{sec:guendel}.  And, in this sense, these components are very
special.  Now we define $\bbox{\epsilon}$ in terms of its
components:
\begin{eqnarray} \bbox{\epsilon} &:=&\frac{1}{4!}\,\bbox{\epsilon}_{ijkl}
\,dx^i\wedge \label{Levi}
dx^j\wedge dx^k\wedge dx^l\;, \\  
& & \qquad \qquad \text{with}\quad
\bbox{\epsilon}_{0123}=1=\text{invariant}\;. \nonumber
\end{eqnarray} 
Consequently $\bbox{\epsilon}$ transforms as an {\em odd 4-form
density of weight $-1$} (see, e.g., \cite[Appendix A]{hehl95}, for
details),  
\begin{equation}\bbox{\epsilon}'=\frac{1}{J}\;
\bbox{\epsilon}=\frac{\text{sgn}{J}}{|J|}\;\bbox{\epsilon}
\;,\label{density}\end{equation} 
where $J=\text{det}\bigl(\partial x^j/\partial {x'}^i\bigr)$ is the
determinant of the Jacobian matrix of the diffeomorphism
$x^i\rightarrow {x'}^i(x^j)$. We are denoting densities by boldface
letters.  

We note in passing that for the manipulation of the
$\bbox{\epsilon}$-basis the following algebraic rules will
turn out to be useful. 
If we take the interior product $\rfloor$ of an arbitrary 
frame $e_\alpha$ with the Levi-Civita $\bbox{\epsilon}$ 4-form
density, then we find a 3-form
$\bbox{\epsilon}_\alpha$; if we contract again, we find a 2-form
$\bbox{\epsilon}_{\alpha\beta}$, etc.:
\begin{mathletters}\label{car}
\begin{eqnarray}
  \bbox{\epsilon}_{\alpha} &:=& e_{\alpha} \rfloor
  \bbox{\epsilon} = \frac{1}{3!}\,
  \bbox{\epsilon}_{\alpha\beta\gamma\delta}\, \vartheta^{\beta} \wedge
  \vartheta^\gamma \wedge \vartheta^{\delta}\;, \\ 
  \bbox{\epsilon} _{\alpha\beta}& := & e_{\beta} \rfloor
  \bbox{\epsilon} _{\alpha} =
  \frac{1}{2!}\,\bbox{\epsilon}_{\alpha\beta\gamma\delta} \, \vartheta
  ^{\gamma}\wedge \vartheta ^{\delta} \;, \\ 
  \bbox{\epsilon}_{\alpha\beta\gamma}& := & e_{\gamma} \rfloor
  \bbox{\epsilon}_{\alpha\beta}=
  \frac{1}{1!}\,\bbox{\epsilon}_{\alpha\beta\gamma\delta}\, \vartheta^\delta\;,
  \\ \bbox{\epsilon}_{\alpha\beta\gamma\delta}\, & \;= & 
  e_\delta\rfloor\bbox{\epsilon}_{\alpha\beta\gamma}=e_{\delta}\rfloor
  e_\gamma\rfloor e_\beta\rfloor
  e_\alpha\rfloor\bbox{\epsilon}\;.
\end{eqnarray}
\end{mathletters}
Here, the coframe $\vartheta^\beta$ is dual to the frame $e_\alpha$,
that is, $e_\alpha \rfloor \vartheta^\beta = \delta_\alpha^\beta$.  The
$(\bbox{\epsilon}\,,\bbox{\epsilon}_{\alpha}\,,
\bbox{\epsilon}_{\alpha\beta}\,,
\bbox{\epsilon}_{\alpha\beta\gamma}\,,
\bbox{\epsilon}_{\alpha\beta\gamma\delta})$ represent a basis
for the odd form densities of weight $-1$.  It is called
$\bbox{\epsilon}$-basis and can be used to define a metric
independent duality operation.  Instead of lowering the rank of the
$\bbox{\epsilon}$'s, we can also increase their rank by exterior
multiplication with the coframe $\vartheta^\mu$:
\begin{mathletters}\label{cons}
\begin{eqnarray}
\vartheta^\mu\wedge\bbox{\epsilon}_{\alpha}& 
  = & +\delta^\mu_{\alpha}\, \bbox{\epsilon}\;, \\ 
  \vartheta^\mu\wedge\bbox{\epsilon}_{\alpha\beta}&
  = & -\delta^\mu_{\alpha }\,
  \bbox{\epsilon}_{\beta}+\delta^\mu_{\beta}\,
  \bbox{\epsilon}_{\alpha }\;,\\ 
  \vartheta^\mu\wedge\bbox{\epsilon}_{\alpha\beta\gamma}&
  = & +\delta^\mu_{\alpha}\, \bbox{\epsilon}_{\beta\gamma}
  -\delta^\mu_{\beta}\,\bbox{\epsilon}_{\alpha\gamma}
  +\delta^\mu_{\gamma}\,\bbox{\epsilon}_{\alpha\beta}\;, \\ 
  \vartheta^\mu\wedge\bbox{\epsilon}_{\alpha\beta\gamma\delta}&
  = &- \delta^\mu_{\alpha}\,
  \bbox{\epsilon}_{\beta\gamma\delta}+\delta^\mu_{\beta}\,
  \bbox{\epsilon}_{\alpha\gamma\delta}-\delta^\mu_\gamma\,
  \bbox{\epsilon}_{\alpha\beta\delta}
  +\delta^\mu_{\delta}\,\bbox{\epsilon}_{\alpha\beta\gamma}\;.
\nonumber\\
\end{eqnarray}
\end{mathletters} 
For the $\widetilde{\epsilon}$ (which is an odd 4-form density of weight 0),
formulae analogous to (\ref{car}),(\ref{cons}) are valid. We have just
to add twiddles to the $\epsilon$'s.

Since $\bbox{\epsilon}$ is an odd density of weight $-1$, we can
split the volume element $\widetilde{\epsilon}$, if we postulate the
existence of an {\em even scalar density} $\bbox{\sigma}$ of
weight $+1$, that is, \begin{equation} \bbox{\sigma}'=
|J|\,\bbox{\sigma} \;.\label{sigma}\end{equation} For our purpose
here\footnote{In reference \cite[Eq.(A.1.33)]{hehl95} we took an
  {\em odd} scalar density instead, which we also denoted by the same
  letter $\sigma$.}, we postulated an {\em even} scalar density,
since the $\widetilde{\epsilon}$ in (\ref{eq:action}) and the Levi-Civita
$\bbox{\epsilon}$ in (\ref{Levi}) are both odd. Then eventually,
equation (\ref{eq:action}) can be rewritten as 
\begin{eqnarray}
W & = & \int\underbrace{L}_{\text{odd 4-f.}} = \int
\nonumber \underbrace{\widetilde{\epsilon}}_{\text{odd
    4-f.}}\underbrace{\widehat{L}}_{\text{scalar}} \\ 
& = & \int \underbrace{\bbox{\epsilon}}_{\text{odd
    4-f.\,density,}\atop\text{weight
$-1$}}\underbrace{\bbox{\sigma}}_{\text{even
scalar density,}\atop\text{weight $+1$}}\quad
\underbrace{\widehat{L}}_{\text{scalar}}\;.\label{rewritten} 
\end{eqnarray}
The scalar density $\bbox{\sigma}$, in contrast to the
Levi-Civita $\bbox{\epsilon}$, must be specified in some way,
before one can actually do physics on the spacetime manifold. It is
here where gravity comes in.

\section{Choices for the scalar density $\bbox{\sigma}$}
\label{sec:sec3}\relax
\subsection{Metric}
It is conventional wisdom to choose the square root of the modulus of
the metric determinant as the scalar density for building up the
volume element:
\begin{equation} \bbox{{}_0\sigma}
:= \sqrt{\left| \det g_{ij}\right|}\;.\end{equation} 
As soon as a metric $g=g_{ij}\,dx^i\otimes dx^j$ is given---the
gravitational potential of general relativity---we can define
$\bbox{{}_0\sigma}$. In conventional integration theory, this is
called the volume measure. We prefer to call it the metric volume
measure and, accordingly, $\eta:=\bbox{{}_0\sigma\,\epsilon}$ the
{\em metric volume element}. Remember that Einstein, in his 1916
review paper of general relativity, see \cite[p.\ 304]{einstein}, only
admitted coordinates such that
$\bbox{{}_0\sigma}\stackrel{*}{=}1$. This amounted to use the
`purely geometrical' volume element $\bbox{\epsilon}$ by
constraining the free choice of the coordinates. It should be noted,
however, that Einstein, in formulating general relativity, did
{\em not} restrict the diffeomorphism invariance in any way, as he
stressed himself (loc.cit.), to what is sometimes called a
volume-preserving diffeomorphism (see \cite{buch88,buch89,We7}). For
him, it was only a convenience in evaluating the diffeomorphism
covariant equations of general relativity.

\subsection{Dilaton field}
Alternatively, we can promote the scalar density to a new fundamental
field of nature, compare also the model developed in \cite[Sec.6]{hehl95}.
The value of such a density $\bbox{{}_1\sigma}$ can be
viewed as a scale factor of the volume element, see also
\cite{buch88,We7}. Thus, from a physical 
point of view, it is interesting to investigate the role of
$\bbox{{}_1\sigma}$ as a scaling parameter which realizes a
scale transformation on a physical system.

Scale transformations as symmetry transformations play an important
part in physics. In particular in the context of cosmological models
and the study of the unification of the four interactions it is common
to start from a scale invariant theory (for the sake of
renormalizability, e.g.) and later derive the scales nowadays observed
by some mechanism like spontaneous symmetry breaking.

Clearly, scale invariance has to be carefully defined---and there are
several possibilities. Well-known scale transformations are the Weyl
transformations $g_{ij}\longrightarrow e^{2\Phi} g_{ij}$ which can be
thought of as a rescaling of the metric.  In conventional theories,
where the volume element is taken as $\sqrt{\left|
\det g_{ij} \right|}$, a Weyl transformation scales the volume
element in a definite way. This is important if Weyl-invariance of an
action integral is considered as it is the case in string
theories. The possibility to take a metric 
independent scalar density $\bbox{{}_1\sigma}$ in place of
$\sqrt{\left| \det g_{ij}\right|}$ in order to build a proper volume
element gives new opportunities to define a concept of scale
invariance. In this context, the field $\bbox{{}_1\sigma}$ is
known as a dilaton field, which becomes non-trivial in the quantum
theory after the conformal (scale) invariance is broken.

For example, a replacement of $\sqrt{\left| \det g_{ij}\right|}$ by
an independent $\bbox{{}_1\sigma}$ will, a priori, `decouple'
the scaling properties of the volume element from the scaling
properties of the (gravitational and matter) Lagrangian. Moreover, the
gravitational field equations which are obtained from varying the
metric, will, in general, be changed. In particular, the metrical
energy momentum current will take a modified form, due to the absence
of the factor $\sqrt{\left| \det g_{ij}\right|}$. Here, the
possibility of introducing the independent field
$\bbox{{}_1\sigma}$ can lead to a modification of this current
and (possibly) related anomalies.\footnote{In this respect the
computations by Buchm\"uller and Dragon \cite{buch89} concerning the
quantization of gravitation in the volume-preserving gauge
$\bbox{{}_0\sigma} \stackrel{*}{=} 1$ should be useful as a
first ansatz.} 
In this sense, employing
$\bbox{{}_1\sigma}$ as independent scale parameter yields a
lot of opportunities for physical creativity. We will come back to
this question elsewhere.

\subsection{Linear connection}
In a pure connection ansatz, we prescribe a linear {connection}
$\Gamma_\alpha{}^\beta = \Gamma_{i\alpha}{}^\beta dx^i$ (but {\em
no} metric!). Define, as usual, the curvature-2-form by 
\begin{equation}
R_\alpha{}^\beta = d\Gamma_\alpha{}^\beta - \Gamma_\alpha{}^\gamma
\wedge \Gamma_\gamma{}^\beta \end{equation}
and the Ricci-1-form by
\begin{equation}
\text{Ric}_\alpha := e_\beta \rfloor R_\alpha{}^\beta = \text{Ric}_{i\alpha}\,
dx^i \;. \end{equation} Then \begin{equation} \bbox{{}_2\sigma} :=
\sqrt{\left| \det \text{Ric}_{ij}\right|} \;, \end{equation} 
with $\text{Ric}_{ij} = \text{Ric}_{i\alpha} e_j{}^\alpha$, is a
viable scalar density, as first suggested by Eddington
\cite[Sec.\ 92]{Eddington}, compare also Schr\"odinger \cite{schr60} 
and the more recent work of Kijowski and collaborators, see
\cite{Kijowski}. Likewise, the corresponding quantity based on the
{\em symmetric} part of the Ricci tensor, 
\begin{equation} \bbox{{}_3\sigma}
:= \sqrt{\left| \det \text{Ric}_{(ij)}\right|} \;, \end{equation} 
also qualifies as a volume measure.  Note that $\bbox{{}_2\sigma}$ and
$\bbox{{}_3\sigma}$ may have singular points in such a theory as
soon as the Ricci tensor or its symmetric part vanish. There seems to
exist no criterion around which would prefer, say,
$\bbox{{}_2\sigma}$, as compared to $\bbox{{}_3\sigma}$.
Lately, such theories have been abandoned.

\subsection{A quartet of scalar fields}
In close analogy to the components $\bbox{\epsilon}_{ijkl}$ of
the Levi-Civita $\bbox{\epsilon}$, we can define the totally
antisymmetric tensor density $\bbox{\epsilon}^{ijkl}$ of
weight $+1$. We put its numerically invariant component
$\bbox{\epsilon}^{0123} = -1$.\label{sec:guendel}

We then can define \cite{Guendel996a}
\begin{eqnarray} \nonumber \bbox{{}_4\sigma} &:= & - 
  \bbox{\epsilon}^{ijkl} \, \left(\partial_i\varphi^{(0)} \right)
  \left(\partial_j\varphi^{(1)} \right) \left(\partial_k\varphi^{(2)} \right)
  \left(\partial_l\varphi^{(3)} \right) \\ & = & - \frac{1}{4!}
  \,\bbox{\epsilon}^{ijkl}\, \epsilon_{ABCD}\, \left(
    \partial_i\varphi^A\right)\left(\partial_j\varphi^B\right)\left(
    \partial_k\varphi^C\right) \left(\partial_l\varphi^D\right) \;,
\nonumber \\
\label{eq:guendel}\end{eqnarray}
where $A,\dots,D$ are indices of interior space.  This definition
yields, for the volume 4-form 
\begin{equation}
\overline{\eta}:=\bbox{{}_4\sigma\,\epsilon}\;,\end{equation} 
the following relations:
\begin{eqnarray}
  \overline{\eta} & = & d\varphi^{(0)} \nonumber
  \wedge d\varphi^{(1)} \wedge d\varphi^{(2)} \wedge d\varphi^{(3)} \\
  & = & \frac{1}{4!}\;{\epsilon}_{ABCD}\;d\varphi^{A} \wedge d\varphi^{B}
  \wedge d\varphi^{C} \wedge d\varphi^{D}\;.
\label{eq:guendel-ext} \end{eqnarray}
If we introduce the abbreviation 
\begin{equation}
\partial_A:=\frac{\partial}{\partial\varphi^A}\;,\label{partial}
\end{equation} 
then the duality of $d\varphi^A$ and $\partial_B$ can be expressed as
follows:
\begin{equation} d\varphi^A\left[\partial_B
\right]=\delta_B^A \;.\end{equation} 
In analogy to the set of equation (\ref{car}), we define the 3-form and the 
2-form
\begin{equation}
\overline{\eta}_A:=\partial_A\rfloor\overline{\eta}\;,\quad
\overline{\eta}_{AB}:=\partial_B\rfloor\overline{\eta}_A\;,\quad\text{etc.}
\label{carbar}\end{equation} 
Explicitly they read
\begin{mathletters}
\label{explicit}
\begin{eqnarray}
\overline{\eta}_A & = & \frac{1}{3!}{\epsilon}_{ABCD}d\varphi^B\wedge
d\varphi^C \wedge d\varphi^D\;, \\ 
\overline{\eta}_{AB} & = & \frac{1}{2!}{\epsilon}_{ABCD}
d\varphi^C \wedge d\varphi^D\;,\quad\text{etc.}
\end{eqnarray}
\end{mathletters}
In analogy to (\ref{cons}) we have 
\begin{equation}
d\varphi^N\wedge\overline{\eta}_A=+\delta^N_A\,\overline{\eta}\;,
\quad d\varphi^N\wedge\overline{\eta}_{AB}=-\delta^N_A\,
\overline{\eta}_B+\delta^N_B\,\overline{\eta}_A\;,
\label{consbar1}\end{equation} 
and so on. We contract (\ref{consbar1}) and find
\begin{equation}
\overline{\eta}=\frac{1}{4}\,d\varphi^N\wedge\overline{\eta}_N\;,\quad
\overline{\eta}_A=\frac{1}{3}\,d\varphi^N\wedge\overline{\eta}_{AN}\;,\quad
\text{etc.}\label{consbar2} \end{equation} 
We differentiate (\ref{consbar2}): 
\begin{equation}
d\,\overline{\eta}=-\frac{1}{4}\,d\varphi^N\wedge
d\,\overline{\eta}_N\;,\quad
d\,\overline{\eta}_A=-\frac{1}{3}\,d\varphi^N\wedge
d\,\overline{\eta}_{AN}\;,\;\; \text{etc.}\label{dconsbar2} \end{equation}

Now, $\overline{\eta}$, as a 4-form, is closed: 
\begin{equation}
d\,\overline{\eta}=0\;.\end{equation} 
Provided $d\varphi^A\neq 0$, we find successively, 
\begin{equation}
d\,\overline{\eta}_A=0\;,\quad 
d\,\overline{\eta}_{AB}=0\;,\quad\text{etc.}
\label{success}
\end{equation} 
Using this information, we can partially integrate (\ref{consbar2})
and can prove that all these forms are not only closed, but also
exact:  
\begin{equation}
\overline{\eta}=d\left[\frac{1}{4}\,\varphi^N\wedge\overline{\eta}_N
\right]\;,\quad \overline{\eta}_A=
d\left[\frac{1}{3}\,\varphi^N\wedge\overline{\eta}_{AN}
\right]\;,\quad \text{etc.}\label{consbar3} \end{equation}

Using (\ref{eq:guendel-ext}) and (\ref{explicit}), we find 
\begin{equation}
\frac{\partial\,\overline{\eta}}{\partial\,{d}\varphi^A}=\overline{\eta}_A
\;,\quad \frac{\partial\,\overline{\eta}_A}{\partial\,{d}\varphi^B}=
\overline{\eta}_{AB}\;,\quad\text{etc.}\label{lagrangemom1}\end{equation}
or, because of (\ref{consbar3}): 
\begin{equation}
d\;\frac{\partial\,\overline{\eta}}{\partial\,{d}\varphi^A}=0\;,\quad
d\;\frac{\partial\,\overline{\eta}_A}{\partial\,{d}\varphi^B}=0
\;,\quad\text{etc.}\label{lagrangemom2}\end{equation} 
Since the corresponding ``forces'' vanish too, as can be seen from
(\ref{eq:guendel-ext}) and  (\ref{explicit}), 
\begin{equation}
\frac{\partial\,\overline{\eta}}{\partial\,\varphi^A}=0\;,\quad
\frac{\partial\,\overline{\eta}_A}{\partial\,\varphi^B}=0
\;,\quad\text{etc.,}\label{lagrangeforce}\end{equation} 
we find an analogous result for the variational derivatives:
\begin{equation} \frac{\delta\, 
  \overline{\eta}}{\delta\, \varphi^A}=0\;,\quad \frac{\delta\,
  \overline{\eta}_A}{\delta\, \varphi^B}=0
\;,\quad\text{etc.}\label{eulerlagrange}\end{equation}
Similarly, we have 
\begin{equation}
\frac{\delta\overline{\eta}}{\delta\vartheta^\alpha}=0\;,\quad
\frac{\delta\overline{\eta}_A}{\delta\vartheta^\alpha}=0\;,\quad\text{etc.}
\label{sim1}
\end{equation}
and
\begin{equation}
\frac{\delta\overline{\eta}}{\delta g_{\alpha\beta}}=0\;,\quad
\frac{\delta\overline{\eta}_A}{\delta g_{\alpha\beta}}=0\;,\quad
\text{etc.}
\label{sim2}
\end{equation}

That the volume element is an exact form is the distinguishing
feature of this ansatz. It is for that reason why the existence of
a quartet of fundamental scalar fields is required instead of only one
scalar field.  Under these circumstances, the volume (\ref{volume}) can be
expressed, via Stokes' theorem, as a 3-dimensional surface integral
which doesn't contribute to the variation of the action functional.

\section{The quartet theory}
\label{sec:sec5}
Using Eq.\ (\ref{rewritten}) and the volume element 
(\ref{eq:guendel-ext}) and denoting the gravitational Lagrangian by
$V=V(g_{\alpha\beta},\vartheta^\alpha,Q_{\alpha\beta},T^\alpha,R_\alpha
{}^\beta)$ and the matter Lagrangian by $L_{\text{m}}=
L_{\text{m}}(g_{\alpha\beta},\vartheta^\alpha,
\Gamma_\alpha{}^\beta,\Psi,d\Psi)$, the action $W$
reads 
\begin{equation} W = \int (V+L_{\text{m}})=\int \overline{\eta}\;
\underbrace{(\widehat{V}+\widehat{L}_{\text{m}})}_{\text{scalar}}=\int
d\,\chi\, (\widehat{V}+\widehat{L}_{\text{m}})\;,
\end{equation} 
see Guendelman and Kaganovich \cite{Guendel996a,Guendel997a,Guendel997b}. 
Note that the 3-form $\chi$, according to (\ref{consbar3}), explicitly
reads $\chi:=\varphi^N\wedge\overline{\eta}_N /4$. If we add a
constant $\lambda$ to the scalar Lagrangian, we find 
\begin{equation}\int d\,\chi\,
(\widehat{V}+\widehat{L}_{\text{m}}+\lambda)=W+\lambda\int d\,\chi\;.
\end{equation} 
Since the 3-dimensional hypersurface integral
$\int_{\partial\,\text{Vol}}\chi$ doesn't contribute to the variation,
the scalar Lagrangian is invariant under the addition of a constant.

Variation with respect to $\varphi^A$ yields the corresponding field
equations 
\begin{equation} \frac{\partial\; (V+L_{\text{m}})}{\partial
\varphi^A} - d\; \frac{\partial\; (V+L_{\text{m}})}{\partial\,
d\varphi^A} = 0 \;.\end{equation} 
Suppose, see \cite{Guendel996a,Guendel997a,Guendel997b}, that
$\widehat{V}$ and $\widehat{L}_{\text{m}}$ do {\em not} depend on the
quartet field at all, 
\begin{equation} 
\frac{\partial\,\widehat{V}}{\partial\,\varphi^A}=0\;,\quad
\frac{\partial\,\widehat{V}}{\partial\, d\varphi^A}=0\;,\qquad
\frac{\partial\,\widehat{L}_{\text{m}}}{\partial\,\varphi^A}=0\;,\quad
\frac{\partial\,\widehat{L}_{\text{m}}}{\partial\, d\varphi^A}=0 
\;,\label{nodepend}\end{equation} 
then the field equations for the quartet field
read 
\begin{equation} (\widehat{V}+\widehat{L}_{\text{m}})\;
\frac{\partial\,\overline{\eta}}{\partial \varphi^A}-d\Bigl[
(\widehat{V}+\widehat{L}_{\text{m}})\,\frac{\partial\;
  \overline{\eta}}{\partial\;d\varphi^A} \Bigr] = 0\;.
\label{eq:field-phi}\end{equation}  
The first term vanishes, since $\overline{\eta}$ does not depend on
$\varphi^A$ explicitly, see (\ref{lagrangeforce}). Then the Leibniz
rule yields 
\begin{eqnarray} d & \Bigl[ ( & \widehat{V} + \widehat{L}_{\text{m}}) \,
  \frac{\partial\; \overline{\eta}}{\partial\,d\varphi^A} \Bigr]  \\
& = & \frac{\partial\; \overline{\eta}}{\partial\,d\varphi^A} \,
d\;(\widehat{V}+\widehat{L}_{\text{m}}) +
(\widehat{V}+\widehat{L}_{\text{m}}) \, 
\underbrace{d\frac{\partial\;\overline{\eta}}{\partial\,d
\varphi^A}}_{\stackrel{(\ref{lagrangemom2})}{=}0} = 0\;.\end{eqnarray}
Provided $\varphi^A\neq 0$ and $d\varphi^A\neq 0$, we
can conclude that 
\begin{equation} d\;(\widehat{V}+\widehat{L}_{\text{m}}) =0\;,
\qquad \text{i.e.,}\qquad \widehat{V}+\widehat{L}_{\text{m}} =
\text{const}\;.
\label{phifieldeq}\end{equation}

The gravitational field equations following from $\delta
g_{\alpha\beta}$ and $\delta \Gamma_\alpha{}^\beta$ are {\em not}
disturbed by the existence of $\varphi^A$. Hence the
usual metric-affine formalism applies in its conventional form (see
\cite{hehl95}, for recent developments cf.\ 
\cite{yuval,yuri,alfredo,garcia}), but the field equation
(\ref{phifieldeq}) for the scalar field quartet $\varphi^A$ has to
be appended. Perhaps surprisingly, it is only one equation since,
in addition to (\ref{sim1}) and (\ref{sim2}), we trivially have
\begin{equation}
\frac{\delta\overline{\eta}}{\delta \Gamma_{\alpha}{}^\beta}=0\;,\quad
\frac{\delta\overline{\eta}_A}{\delta \Gamma_\alpha{}^\beta}=0\;,\quad
\text{etc.}
\label{sim3}
\end{equation}

\section{Conclusion}
We can reproduce the essential features of the Guendelman-Kaganovich
theory without the necessity to specify the gravitational first-order
Lagrangian other than by the property (\ref{nodepend}).

\acknowledgments
We would like to thank Eduardo Guendelman and Alex Kaganovich for
useful comments and remarks.
This research was supported by CONACyT, grants No.\ 3544--E9311,
No.\ 3898P--E9608, and by the joint German--Mexican project
DLR(BMBF)-CONACyT MXI 6.B0A.6A.

\appendix
\section*{Covariant exterior derivative of
the $\bbox{\epsilon}$-basis}
\label{sec:sec4} 
For computations with volume elements, it is convenient to
introduce the differentials of the $\bbox{\epsilon}$- and the
$\widetilde{\epsilon}$-basis. As soon as a linear connection 1-form
$\Gamma_\alpha{}^\beta=\Gamma_{i\alpha}{}^\beta dx^i$ is given --
there is no need of a metric for that -- we find by covariant exterior
differentiation of (\ref{car}),
\begin{mathletters}
\begin{eqnarray}
  D\bbox{\epsilon} _{\alpha}& = & T^\mu \wedge
  \bbox{\epsilon} _{\alpha\mu }\;,\\ 
  D\bbox{\epsilon} _{\alpha\beta}& = & T^\mu \wedge
  \bbox{\epsilon} _{\alpha\beta\mu}\;,\\ 
  D\bbox{\epsilon}_{\alpha\beta\gamma}&
  = & T^\mu\wedge\bbox{\epsilon}_{\alpha\beta\gamma\mu} \;,\\ 
  D\bbox{\epsilon}_{\alpha\beta\gamma\delta} & = & 0 \;.
\end{eqnarray}
\end{mathletters}
Here $T^\alpha:= D\vartheta^\alpha$ is the torsion 2-form. For the
$\bbox{\sigma}$-modified $\widetilde{\epsilon}$-basis, the
computations run on the same track,
\begin{mathletters}\label{epsilontilde}
\begin{eqnarray}
  D\widetilde{\epsilon}_{\alpha}&
  = & \frac{D\bbox{\sigma}}{\bbox{\sigma}}\wedge
  \widetilde{\epsilon}_{\alpha}+ T^\mu \wedge \widetilde{\epsilon}_{\alpha\mu
    }\;,\\ 
  D\widetilde{\epsilon}_{\alpha\beta}&
  = & \frac{D\bbox{\sigma}}{\bbox{\sigma}}\wedge \widetilde{\epsilon}
  _{\alpha\beta} +T^\mu \wedge
  \widetilde{\epsilon}_{\alpha\beta\mu}\;,\\ 
  D\widetilde{\epsilon}_{\alpha\beta\gamma}&
  = & \frac{D\bbox{\sigma}}{\bbox{\sigma}}\wedge
  \widetilde{\epsilon}_{\alpha\beta\gamma}+
  T^\mu\wedge\widetilde{\epsilon}_{\alpha\beta\gamma\mu}\;, \\ 
  D\widetilde{\epsilon}_{\alpha\beta\gamma\delta}&
  = & \frac{D\bbox{\sigma}}{\bbox{\sigma}}\wedge\widetilde{\epsilon}
  _{\alpha\beta\gamma\delta}\;.
\end{eqnarray}
\end{mathletters}
the advantage being that this basis is composed of forms of weight 0,
i.e., not of densities.

If a metric $g$ is given additionally, then we can take
$\bbox{{}_0\sigma}$ as scalar density and find generally, 
\begin{equation}
\frac{D\bbox{{}_0\sigma}}{\bbox{{}_0\sigma}}=-2\,Q\;,\label{null}
\end{equation} 
with the Weyl covector $Q:=Q_\gamma{}^\gamma/4$ and the nonmetricity
1-form $Q_{\alpha\beta}:=-Dg_{\alpha\beta}$. It is then simple to
rewrite (\ref{epsilontilde}) in terms of the metric volume 
element $\eta:=\bbox{{}_0\sigma\,\epsilon}$:
\begin{mathletters}\label{etavol}
\begin{eqnarray}
  D\eta_{\alpha}& = & -2Q\wedge \eta _{\alpha}+ T^\mu \wedge \eta
  _{\alpha\mu }\;,\\ 
  D\eta _{\alpha\beta}& = & -2Q\wedge \eta
  _{\alpha\beta} +T^\mu \wedge \eta _{\alpha\beta\mu}\;,\\ 
  D\eta _{\alpha\beta\gamma}& = & -2Q\wedge \eta _{\alpha\beta\gamma}+
  T^\mu\wedge\eta_{\alpha\beta\gamma\mu}\;, \\ 
  D\eta_{\alpha\beta\gamma\delta}& = & -2Q\wedge\eta
  _{\alpha\beta\gamma\delta}\;.
\end{eqnarray}
\end{mathletters}
These equations turn out to be very helpful in conventional
applications. However, in the quartet theory, we have to forget
(\ref{etavol}) and to take recourse to $\bbox{_4\sigma}$. 

Thus, analogously to (\ref{epsilontilde}), we have
\begin{mathletters}\label{etabartilde}
\begin{eqnarray}
  D\overline{\eta}_{\alpha}&
  = & \frac{D\bbox{_4\sigma}}{\bbox{_4\sigma}}\wedge
  \overline{\eta}_{\alpha}+ T^\mu \wedge \overline{\eta}_{\alpha\mu
    }\;,\\ 
  D\overline{\eta}_{\alpha\beta}&
  = & \frac{D\bbox{_4\sigma}}{\bbox{_4\sigma}}\wedge \overline{\eta}
  _{\alpha\beta} +T^\mu \wedge
  \overline{\eta}_{\alpha\beta\mu}\;,\\ 
  D\overline{\eta}_{\alpha\beta\gamma}&
  = & \frac{D\bbox{_4\sigma}}{\bbox{_4\sigma}}\wedge
  \overline{\eta}_{\alpha\beta\gamma}+
  T^\mu\wedge\overline{\eta}_{\alpha\beta\gamma\mu}\;, \\ 
  D\overline{\eta}_{\alpha\beta\gamma\delta}&
  = & \frac{D\bbox{_4\sigma}}{\bbox{_4\sigma}}\wedge\overline{\eta}
  _{\alpha\beta\gamma\delta}\;.
\end{eqnarray}
\end{mathletters}
We find 
\begin{equation} \frac{D\bbox{{}_4\sigma}}{\bbox{{}_4\sigma}}=
\frac{d\bbox{{}_4\sigma}}{\bbox{{}_4\sigma}}-
\Gamma_\alpha{}^\alpha \;, \label{relvol}\end{equation} 
but we were not able to find a more compact expression for
(\ref{relvol}).  However, provided a metric is present besides the
scalar quartet and the connection, we can rewrite (\ref{relvol}) as
follows: 
\begin{equation}\frac{D\bbox{{}_4\sigma}}{\bbox{{}_4\sigma}}=
\frac{d\bbox{{}_4\sigma}}{
  \bbox{{}_4\sigma}}- 2Q+d\ln\sqrt
{|\text{det}\,g_{\alpha\beta}|}\;. \end{equation}


\end{multicols}
\end{document}